# *algoXSSF:* Detection and analysis of cross-site request forgery (XSRF) and cross-site scripting (XSS) attacks via Machine learning algorithms

Naresh Kshetri
*School of Business & Technology*
*Emporia State University*
Emporia, KS, USA
nkshetri@emporia.edu

Dilip Kumar
*School of Management*
*Indian Institute of Technology*
Mandi, India
d23233@students.iitmandi.ac.in

James Hutson
*Dept. of Art History & Culture*
*Lindenwood University*
Saint Charles, MO, USA
jhutson@lindenwood.edu

Navneet Kaur
*Dept. of Computer Science*
*University of Missouri-St. Louis*
St. Louis, MO, USA
nk62v@umsystem.edu

Omar Faruq Osama
*Dept of Sys Sci & Ind Eng.*
*Binghamton University, SUNY*
Binghamton, NY, USA
osamaomarfaruq@gmail.com

*Abstract*— The global rise of online users and online devices has ultimately given rise to the global internet population apart from several cybercrimes and cyberattacks. The combination of emerging new technology and powerful algorithms (of Artificial Intelligence, Deep Learning, and Machine Learning) is needed to counter defense web security including attacks on several search engines and websites. The unprecedented increase rate of cybercrime and website attacks urged for new technology consideration to protect data and information online. There have been recent and continuous cyberattacks on websites, web domains with ongoing data breaches including - GitHub account hack, data leaks on Twitter, malware in WordPress plugins, vulnerability in Tomcat server to name just a few. We have investigated with an in-depth study apart from the detection and analysis of two major cyberattacks (although there are many more types): cross-site request forgery (XSRF) and cross-site scripting (XSS) attacks. The easy identification of cyber trends and patterns with continuous improvement is possible within the edge of machine learning and AI algorithms. The use of machine learning algorithms would be extremely helpful to counter (apart from detection) the XSRF and XSS attacks. We have developed the algorithm and cyber defense framework - algoXSSF with machine learning algorithms embedded to combat malicious attacks (including Man-in-the-Middle attacks) on websites for detection and analysis.

## I. INTRODUCTION

The contemporary digital landscape is characterized by a significant proliferation of online activities, encompassing a vast array of operations ranging from personal communications to complex financial transactions. This surge in digital interactivity has been facilitated by the global expansion of internet users and devices, thereby engendering an extensive, interconnected global internet community [1]. However, this digital evolution has been paralleled by an escalation in cyber threats, manifesting in various forms including cybercrimes and sophisticated cyberattacks [2]. The prevalence of these threats underscores the urgent necessity for robust cybersecurity measures. In this context, the advent of emerging technologies, particularly those within the realms of artificial intelligence (AI), deep learning, and machine learning (ML), presents a promising avenue for enhancing web security [3]. As [4] note, these advanced technologies possess the potential to significantly augment the capabilities of cybersecurity systems, thereby providing a formidable defense against a myriad of cyber threats.

Notwithstanding the advancements in cybersecurity measures, the digital domain continues to witness an alarming rate of cybercrimes and website attacks. Recent incidents such as the GitHub account hack, data leaks on Twitter, malware intrusions in WordPress plugins, and vulnerabilities in the Tomcat server exemplify the severity and diversity of these threats [5][6][7][8]. Among the myriad of cyber threats, is cryptojacking & ransomware threats, apart from cross-site request forgery (XSRF) and cross-site scripting (XSS) have emerged as two of the most prevalent and damaging forms of attacks [9][10]. These attacks exploit the vulnerabilities inherent in web applications and browsers, posing a significant risk to both individual users and organizations. The detection and analysis of XSRF and XSS attacks are therefore imperative in the contemporary cybersecurity landscape [11]. The development of effective methods for identifying and mitigating these threats is crucial for safeguarding data and ensuring the integrity of online platforms.

In the pursuit of developing 'algoXSSF', a novel algorithm and cyber defense model, this study adopts a methodological approach grounded in the principles of machine learning and cybersecurity. The methodology is designed to enable the detection and analysis of CSRF and XSS attacks, which are among the most insidious threats in the digital domain [12]. This approach underscores the study's commitment to enhancing web security through the integration of advanced technological solutions. The foundation of the methodology lies in the comprehensive collection and preprocessing of data. This involves gathering a diverse array of datasets that encapsulate instances of CSRF and XSS attacks, sourced from various web applications and platforms [13]. The preprocessing stage is critical in refining this data, as [14] point out, ensuring its suitability for the subsequent analytical processes, and involves cleaning, normalizing, and structuring the data to facilitate effective machine learning analysis.

Upon the completion of data preprocessing, the study progresses to the development and training of machine learning models. These models are meticulously crafted to identify patterns and anomalies indicative of XSRF and XSS attacks. The training process utilizes the preprocessed data, as outlined by [15], enabling the models to learn from real-world examples of these cyber threats. The models are subjected to rigorous training regimes, encompassing a range of scenarios to ensure their robustness and accuracy in detecting potential attacks. In addition to model training, the study also emphasizes the importance of continuous testing and validation. This phase involves subjecting the models to various test cases, including known attack vectors and novel scenarios. The objective is to assess the efficacy of the models in accurately identifying and analyzing XSRF and XSS attacks under different conditions. The testing phase is crucial in fine-tuning the models, ensuring their reliability and effectiveness in real-world applications.

Finally, the deployment of 'algoXSSF' represents the culmination of the methodology. This phase involves integrating the trained and tested models into a cyber defense framework. The deployment is designed to be seamless, ensuring that the novel algorithm can operate effectively within existing web security infrastructures. The model functions by continuously monitoring web traffic and application activities, utilizing its machine learning capabilities to detect and analyze any instances of XSRF and XSS attacks. The intended results of this comprehensive methodology are multifaceted. Primarily, 'algoXSSF' aims to significantly enhance the detection and analysis of such cybersecurity attacks, providing a higher level of security for web applications and platforms. Additionally, the model is expected to contribute to the broader field of cybersecurity by offering insights and learnings that can inform future developments in cyber defense strategies. Ultimately, the deployment of this cyber defense model aspires to establish a new benchmark in the realm of cybersecurity, combining the power of machine learning with advanced cyber defense techniques to protect against the evolving landscape of cyber threats.

## II. BACKGROUND STUDY

First, The literature on cross-site scripting (XSS) and cross-site request forgery (XSRF) is extensive and multifaceted, reflecting the dynamic and evolving nature of these cyber threats. These two prevalent cybercrime techniques pose significant threats to web security. Cross-site scripting involves the injection and execution of malicious scripts within a user's browser, exploiting the trust relationship between the web application and the victim. This can lead to data theft, session hijacking, and malware installation [16]. [17] had previously reinforced this notion, emphasizing XSS as a major threat to web applications, necessitating robust mitigation efforts. [18] further discuss the commonality of XSS vulnerabilities in web applications, suggesting that validating user input through filtering and escaping can effectively prevent these attacks. Conversely, CSRF tricks victims into performing actions on a trusted website without their knowledge, exploiting authenticated sessions and often leveraging social engineering [19]. [20] proposes a light-weight CSRF prevention method, demonstrating the importance of distinguishing between malicious and harmless requests.

This review synthesizes key findings from recent studies, offering insights into the progression, detection, and mitigation of XSS and XSRF attacks. [21] initially proposed a client-side system for automatically detecting XSS vulnerabilities, thereby protecting users and alerting web servers of potential threats. [22] describe a passive detection system for identifying successful XSS attacks, highlighting its efficacy with zero false negatives and an excellent false positive rate. [23], for example, provide a comprehensive overview of the current state of XSS vulnerabilities, highlighting their persistent threat in web applications. Their systematic literature review delineates the evolution of XSS attacks, underscoring their adaptability in the face of changing digital environments. Similarly, [24] explore the evolution of XSRF attacks, focusing on their impact on contemporary web browsers. The study illuminates the sophisticated nature of XSRF threats and their ability to exploit browser vulnerabilities. [25] also contributes to this discourse by elucidating the mechanisms of CSRF attacks, revealing how these can be executed surreptitiously, often without user awareness or intervention.

On the other hand, [26] present a survey on XSS web-attack and defense mechanisms. Their work emphasizes the ongoing challenge of XSS vulnerabilities in popular websites and explores various approaches to mitigate these threats. In addition, [27] offer a detailed classification of XSS attack strategies and corresponding defense mechanisms. Their work provides a comprehensive overview of the state-of-the-art in XSS attack detection and prevention, offering valuable insights for both researchers and practitioners in the field. Additionally, the integration of AI and ML techniques in cybersecurity represents a burgeoning area of research. [15] discuss the trend of using traditional methods to mitigate XSS attacks compared to those utilizing AI techniques. Their survey highlights a growing inclination towards leveraging AI and ML for enhancing cybersecurity measures, particularly in the context of XSS and XSRF threats. In all, these studies reveal a consistent emphasis on the evolving nature of XSS and XSRF attacks and the corresponding need for innovative and effective detection and mitigation strategies. The integration of AI and ML in cybersecurity emerges as a promising area of focus, offering potential advancements in combating these pervasive cyber threats.

## III. XSRF Attacks

Cross-site request forgery (CSRF), alternatively referred to as XSRF or Sea Surf, denotes a form of cyber assault when an adversary deceives a target into executing unintended operations. The Cross-Site Request Forgery (CSRF) is an online assault that is rather straightforward to comprehend, but has continually remained a prominent concern in web security since its first identification in the early 2000s. This practice is commonly executed through the creation of a deceptive request that mimics the origin of a reliable entity, such as a financial institution or an email service provider. In a Cross-Site Request Forgery (CSRF) attack, an illicit website compels the web browser to execute authorized and security-sensitive actions on a specific web application using cross-site requests, bypassing the need for user interaction. The accomplishment of this task may be achieved by using conventional HTML elements and JavaScript, so rendering Cross-Site Request Forgery (CSRF) efforts easily executable. Consequently, web developers that prioritize security must undertake the implementation of measures to counteract malevolent cross-site requests that exploit authentication [28].

Cross-Site Request Forgery (CSRF) attacks possess the capability to illicitly acquire personal information, manipulate account configurations, and perpetrate fraudulent transactions. They are frequently employed alongside other forms of assaults, such as phishing, with the objective of unauthorized entry into the victim's account. CSRF attacks are executed by capitalizing on the inherent vulnerabilities in the manner in which websites manage cookies [29]. Upon a user's login to a web page, the browser proceeds to store a cookie upon the user's computer, thereby establishing an identification mechanism between the user and the website. The purpose of this cookie is twofold: firstly, to maintain the user's logged-in status, and secondly, to verify the user's authorization for specific actions. CSRF attacks can be employed by malicious actors to fabricate requests that simulate origin from the user's web browser. These requests have the potential to execute actions that are not in alignment with the users' intentions, such as initiating a transfer of funds from the user's bank account or altering the password associated with their email account. CSRF attacks can exert a substantial influence on both individuals and businesses. CSRF attacks have the potential to result in adverse consequences for individuals, including money losses, identity theft, and various forms of harm. CSRF attacks have the potential to cause reputational harm, compromise data security, and result in financial ramifications for enterprises [28].

An illustration of the four phases in a cross-site request forgery attack is shown below:

- A falsified request is created by an attacker, and when it is executed, it moves the money from a certain bank into the attacker's account.
- The falsified request is embedded by the attacker into a hyperlink, which is then shared via mass emails and websites.
- After clicking on an email or website link sent by the attacker, the victim requests money transfer from the bank.
- After receiving the request, the bank server considers it valid because the victim has the necessary authorization and transfers the money.

Although CSRF attacks take different approaches, they usually have the following traits:

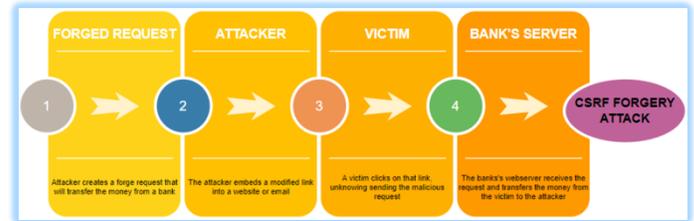

Fig. 1. CSRF forgery attack steps

- They take advantage of websites that depend on user identities.
- They deceive the user's browser into submitting requests for HTTP to the desired website.
- They use HTTP requests with unintended consequences and lack the necessary CSRF safeguards.

The susceptibility of distinct HTTP verbs to cross-site scripting attacks varies, leading to a range of protective measures. This results from the disparate ways in which web browsers process the verbs.

Embedded parameters in HTTP GET requests, such those found inside image tags, are manipulable and vulnerable to attack. For a correctly constructed web application or other resource, GET requests often do not affect state, rendering them useless as targets of cross-site request forgeries (CSRF). Because the state is changed via HTTP POST, more security is required. Web browsers use cross-origin resource sharing (CORS), which includes the cross-origin security policy, and the same origin policy (SOP) as security mechanisms to do this. By restricting a request's or webpage's capacity to communicate with a different origin, the combination of these technologies helps avoid CSRF attacks, among other threats.

In order to mitigate numerous cross-site attacks, other HTTP verbs like PUT and DELETE can only be used with SOP and CORS. Although it's rare, certain websites may specifically turn off these security features, and you can also turn them off directly from within a web browser.

### A. Methods for Mitigating Cross-Site Request Forgery (CSRF) Attacks

Using Anti-CSRF tokens in one of two ways is the most popular approach for thwarting CSRF attacks. The basic idea is the same, even though the token implementations differ slightly: an attacker is less likely to be able to launch an attack without making an extremely improbable guess if they create and compare a randomly generated token string.

- Synchronizer tokens: There exist various strategies to mitigate Cross-Site Request Forgery (CSRF) threats. One widely employed approach involves the utilization of a mechanism known as synchronizer tokens. Synchronizer tokens refer to distinct quantities that are generated by a website and subsequently transmitted to the user's browser. When a web browser initiates a request to a website, it is required to include the synchronizer token. Subsequently, the website will verify the validity of the synchronizer token. In the event that the synchronizer token is deemed invalid, the request will be declined [29].

For example: A random token is embedded into the form when a user accesses a web page, such as the bank's webpage, that facilitates financial transfers. The random token is returned when the user submits the form, allowing the bank to verify if the two tokens match. Transferring takes place if the tokens match. The random token value generated on the web page is inaccessible to the attacker, and even if they were to request the page, they would be unable to read the response due to the same origin policy. This technique obviously has the drawback of making it more difficult for the server to verify if tokens are genuine for each request. If a user has many browser windows open or in other circumstances where the request is being made by other software, it may also cause problems. Part of this challenge may be circumvented by extending the token's scope to include sessions rather than requests.

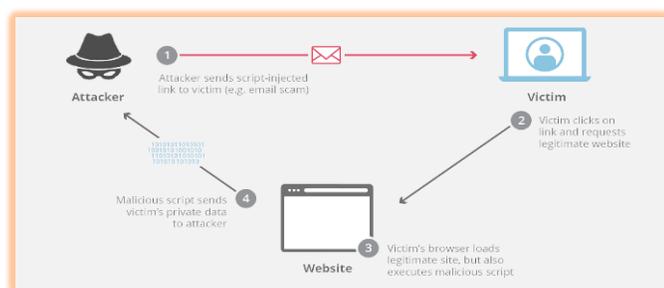

Fig. 2. Script-injected link by attacker

- Cookie-to-header token: An alternative approach is sending a cookie containing a random token to the visitor's browser. The HTTP header that is delivered with each request contains the value of the token that is read by JavaScript running on the client side and copied into it. The value in the header can be checked by the server to make sure the user is sending a legitimate request. A successful assault will be lessened by the failure of any more cases. Users can aid in the prevention of certain CSRF attacks by utilizing custom rules via a web application firewall (WAF) [30].

## IV. XSS Attacks

Cross-Site Scripting (XSS) is an injection technique in which malicious scripts are inserted into reputable websites. An attacker can influence how users interact with a susceptible program through a security flaw known as cross-site scripting (XSS). The identical origin policy, which essentially divides various websites from one another, may be gotten around by an attacker [31]. Through cross-site scripting vulnerabilities, an attacker may often assume the identity of the target user, conduct any operations that the user can accomplish, and access any data that the user has access to. The attacker may be able to take complete control of all the features and data in the program if the target user has elevated access to it. An XSS attack occurs when an online application injects malicious code, often in the form of browser-side script, into another end user [32]. The vulnerabilities that enable the success of this attack are quite prevalent and may arise in any online application that fails to validate or encode user input. According to a survey conducted by Symantic, over 50% of websites are susceptible to XSS attacks. The targets of these attacks were the prominent internet companies of the time, including Twitter, Myspace, Orkut, Facebook, and YouTube. Thus, the term "Cross-Site" Scripting is derived [33].

A cross-site scripting attack (XSS attack) involves the malicious insertion of hazardous or destructive scripts into the source code of trusted software or websites. An XSS attack frequently begins by luring a user into clicking on a malicious hyperlink. Insufficient data sanitation in the app or website allows a malicious link to execute the attacker's covert code on the user's PC. The assailant will thereby get access to the user's data. Cross-site scripting vulnerabilities give an attacker the capability to impersonate the user and get unauthorized access to all of the user's data. If the target user had privileged access to the app or web page, the attacker would have complete control over its functionality and data. Users are routed to a website that is susceptible to cross-site scripting assaults since it includes malicious JavaScript. Once the victim's browser executes the malicious code, the attacker has full control over the use of the capability.

Keep your text and graphic files separate until after the text has been formatted and styled. Do not use hard tabs, and limit use of hard returns to only one return at the end of a paragraph. Do not add any kind of pagination anywhere in the paper. Do not number text heads-the template will do that for you.

### A. Categories of Cross-Site Scripting (XSS)

Cross-Site Scripting (XSS) attacks may result in a range of problems for end users, ranging from minor inconveniences to full compromise of their accounts. The most severe XSS attacks render a user's data susceptible, granting a hacker unauthorized access to the user's identity and account. Additional malicious actions might potentially compromise end-user files, deploy Trojan horse software, redirect users to other websites or web pages, and change how the data is received by the user.

• XSS stored: This happens once the malicious payload becomes stored in a database. If there is no output encoding, the requested data is rendered to other users.

• XSS Reflection: It refers to a security vulnerability where an attacker is able to inject malicious code into a website, which is then executed by the victim's browser. When a web application transmits text supplied by a malicious actor to a user's computer, the browser recognizes and runs a portion of the content as executable code. This is often referred to as an XSS reflection attack. The payload is able to return due to the absence of server-side output encoding.

• Cross-site scripting via the Document Object Model (DOM): It refers to an attack when an adversary injects a script deeper into a server's response. The attacker may manipulate the contents of the Document Object Model (DOM) to generate a harmful URL. The attacker uses this URL to manipulate the victim into accessing it under pretenses. If the user clicks the link, the attacker has access to the user's live session data. By clicking the link, the attacker has unauthorized access to the user's active session information, keystrokes, as well as additional data. DOM-based XSS attacks differ from cached XSS and reflected XSS attacks in that they only target the client browser without any data being sent back to the server. Undoubtedly, an XSS attack occurs when the assailant injects malicious code into the page. Conversely, when a person visits the same page, they can be asked to supply information or, in the worst scenario, they might unknowingly provide sensitive data, such as a password, to a fraudulent website. Without a doubt, the lack of user input verification is a major factor in XSS assaults.

• Non-persistent cross-site scripting (XSS) assault or Reflected XSS: This kind of attack is known as a reflective attack since it entails the web server mirroring its action in response to a user's request for a service, such as search results, a replicated message from the server, or any other response that contains part or all of the data provided to the server. In these attacks, attackers send the link crafted with XSS to get access to the private data in the web application. In this attack, the browser-based application promptly gives the information without making it alright for the program. In this, Information isn't put away on the web server. This attacker tries to Steal cookies and redirect to completely different websites.

• Persistent Cross-Site Scripting (XSS) Attack: In contrast to the non-persistent attack that only manifests as an outcome, this assault actively engages with web pages. Moreover, this assault employs an injection script that will inevitably impact the server's databases in many ways, including comment fields, logs, forums, as well as comparable components. The victim desires to retrieve the previously stored information, which is very probable to include an inserted script [34]. Attackers inject the malicious executable scripts to the browser-based application which are saved in the terms of a record in the database or data log into a log file on target server or webserver. this injected script, can then access session data and cookies of users to perform events on behalf of the client. JavaScript stored into the computer system usually DB. whoever loads over into the site will load the hacker JavaScript as well.

XSS attack is an attack in which an attacker injects malicious executable substance into the code of a trusted application or weak site at client end. To communicate with the web applications, we normally use programs at the client-side and generally speaking, it is the program which assists us with interfacing with web applications, it is the browser. In an XSS attack, we inject malicious code onto the internet browser to cause the web application to accomplish something which in a perfect world shouldn't do. In this case with the help of a web browser an attacker injects malicious script. At the point when the victim visits the site or web server, this malicious program is executed. This attack (as well as fake news detection via machine learning techniques) is most of part used to steal sensitive data like cookies, session tokens and might be other confidential data, might be in the event that we were passing our username or secret key and utilizing this malicious content or using cross-site scripting those data can be taken from the internet browser or the web server [35] [36].

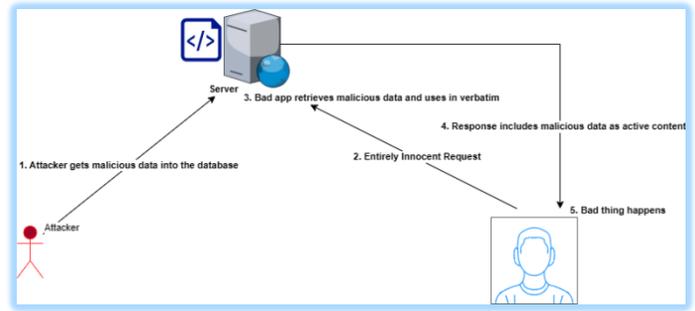

Fig. 3. Stored XSS attack dataflow

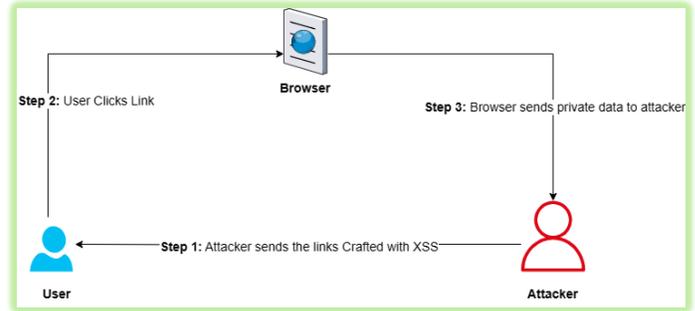

Fig. 4. Reflected XSS attack dataflow

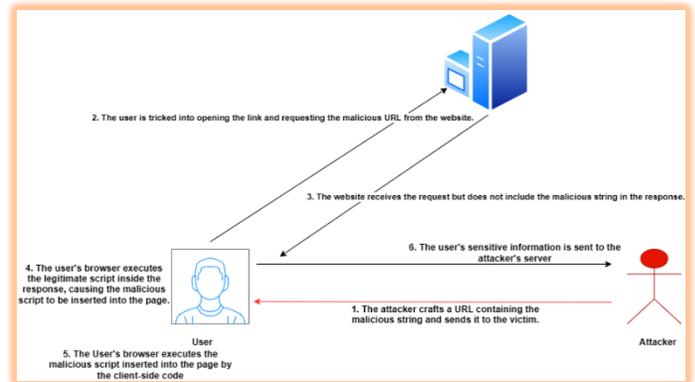

Fig. 5. DOM-based XSS attack dataflow

## V. MACHINE LEARNING ALGORITHMS

The deployment of ML algorithms in detecting and mitigating XSS and CSRF attacks has seen significant advancement in recent years. The advent of ML in the realm of cybersecurity, particularly for the detection and mitigation of such attacks, represents a significant shift towards more dynamic and effective defense mechanisms. A comprehensive analysis of ML and DL approaches for XSS attack detection has been conducted, focusing on various aspects such as domain areas, data preprocessing, feature extraction, feature selection, and dimensionality reduction. This analysis also considers data imbalance, performance metrics, datasets, and data types, providing a holistic view of the current ML/DL strategies in this area [3]. Recent studies have also highlighted the potential of ML algorithms in identifying and countering these threats with promising results. This section synthesizes key findings from the literature, grouped thematically to provide a comprehensive understanding of the current state of research in this field. In fact, a range of ML algorithms have been explored for detecting XSS attacks with promising results.

[37] also aimed to utilize various ML algorithms for detecting XSS attacks. Their research not only applied ML techniques for detection but also compared the performance of these algorithms in their ability to identify XSS attacks effectively. [38] [39] also demonstrated that a combination of blockchain technology, SVM, KNN, and Naïve Bayes algorithms with the n-gram method achieved an accuracy of 98% & improve security in AI based healthcare systems. This finding was complemented by the work of [40], who compared the performance of various algorithms including the support vector method, decision tree, Naive Bayes classifier, and logistic regression. Similarly, [41] focused on SVM, KNN, random forest, and logistic regression, with the random forest classifier exhibiting the highest accuracy. [42] further improved these results, achieving 99.92% accuracy with the AdaBoost classifier. These studies collectively underline potential of ML in effectively detecting XSS attacks. Finally, an experiment described in a study led to model creation for detecting XSS attacks using ML. Model considered various ML algorithms, including support vector method, decision tree, Naive Bayes classifier, and Logistic Regression, demonstrating the diverse tools range available within ML for tackling XSS threats & need for cyber awareness of computer and cyber ethics [43] [44].

In context of CSRF attack detection, ML solutions have also been presented. For example, [45] introduced a ML approach for the black-box detection of CSRF vulnerabilities, highlighting the adaptability of ML algorithms in identifying such security threats. Represents a significant advancement in detection of CSRF attacks, indicating the growing versatility of ML techniques in cybersecurity. Recent research has also focused on integrating more advanced ML techniques. a few explorations are examined that have handled web assaults like XSS and CSRF, utilizing regular ML, improved deep learning techniques.

[46] developed an XSS detection model based on LSTM-Attention, achieving a precision rate of 99.3%, recall rate of 98.2%. Additionally, [47] improved XSS attack detection by combining CNN with LSTM, achieving over 99.4% accuracy in predicting XSS attacks. [48] proposed the framework to detect, monitor and provide prevention methods of DDoS attacks using ML algorithms. Exhibition of 4 most widely utilized algorithms was analyzed with respect to Recall, Accuracy, 1 score, Precision & an analysis tool OWASP, ZAP and Weka is used. [49] state that web applications are especially difficult to break down, because of their variety and the far/wide reception of custom programming rehearses. ML with blockchain technology applications for cyber defense is in this way exceptionally accommodating in web setting since it can exploit physically named information to uncover human comprehension of browser semantics to computerized exam apparatuses [50].

**Steps Involved in Building a M/L Application:**
**Step 1:** Collect data
**Step 2:** Prepare the input data – cleanse, format etc.
**Step 3:** Analyze the input data – Plotting, finding features etc.
**Step 4:** Train algorithm – Run algorithm on training data
**Step 5:** Test algorithm – see to what extent it works on test data
**Step 6:** Use it – If testing successful, use it with new data.

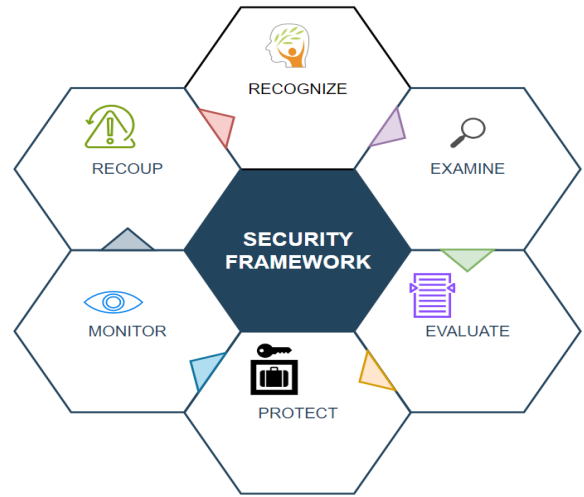

Fig. 6. algoXSSF security framework

VI. DETECTION OF ATTACKS VIA ALGOXSSF

*A. Recognize:* Identify weaknesses in systems, networks, or processes, Analyze project plans, requirements, and other documents to identify potential sources of risk.

*B. Examine:* Prioritize vulnerabilities based on their severity and the potential risk they pose, identification and documentation of vulnerabilities, such as software bugs, configuration errors, or gaps in security controls. Analyze network traffic to identify patterns and anomalies.

*C. Evaluate:* Evaluate the likelihood & impact of each risk, assessing the severity & potential impact of each vulnerability.

*D. Protect:* Secure the wireless access points and networks, install, and activate the software, setup web and email filters, train the employees about the security features. Guard the information timely and effectively.

*E. Monitor:* Maintain and monitor the logs, assess the web programs/ applications for potential security weaknesses and risks, Review logs generated by web servers, applications, and databases., Conduct regular vulnerability assessments to identify and prioritize potential security weaknesses, Perform periodic penetration testing. Monitor for new vulnerabilities, changes in network traffic, and updates to third-party components. Establish continuous monitoring processes to track changes in the web application environment.

*F. Recoup:* Make improvements to the existing plans/ procedures/technologies to combat the attacks in future. Revise the processes, Regularly scan, and monitor systems for new vulnerabilities and apply updates and patches as needed. Develop strategies to reduce or eliminate risks, Define rules and signatures to identify known attack patterns and behaviors, Conduct regular drills to ensure an effective and timely response to security events.

**Algorithm (proposed) for algoXSSF via Machine Learning**
1: Input - $Data_{initial}$
2: Create REC ([Recognize set])
3: Compute $Risk_{BK} = N/REC$
4: **for** $k = 1$ to $log\ N$ **do**
5:     - Transmit using $Data_{next}$ for REC
6:     - Observe $|Risk_{BK}|$ pattern with error, *err*

```
7:         - Detect whether [ATTK, VULN]
8:         - Recoup Data_initial using (4)
9:         - XSRF_d += 1, XSS_d += 1
10: end for
11: Return: flag where XSRF_d = true, XSS_d = true, err
12. Analyze XSRF_d, XSS_d, err values
```

## VII. ANALYSIS AND DISCUSSION

The detection and analysis of Cross-Site Request Forgery (CSRF or XSRF) and Cross-Site Scripting (XSS) attacks through machine learning algorithms represent a proactive and automated approach to safeguarding web applications from malicious exploitation. This methodology involves the utilization of machine learning algorithms that autonomously learn and recognize patterns associated with malicious activities, contributing to the early identification and mitigation of CSRF and XSS threats. In the initial stages, a diverse dataset is collected, encompassing both benign and malicious instances of web requests. Feature extraction becomes a critical step, where relevant attributes such as request headers, tokens, and JavaScript code are identified. These features serve as the basis for the machine learning algorithms to distinguish between normal and malicious activities associated with CSRF and XSS attacks. Various machine learning algorithms can be employed, including supervised learning classifiers like SVM, Decision Trees, or Neural Networks Algorithm. The model is learned on the labeled dataset, allowing it to learn and generalize patterns associated with both benign and malicious instances. Evaluation metrics, like recall, precision, and F1 score, are then used to evaluate the model's accuracy and effectiveness. Feature importance analysis provides insights into the critical characteristics that contribute to the model's decision-making process, facilitating a better understanding of the detected attacks. Regular updates and periodic retraining with new data are essential to ensure the model remains adaptive to emerging attack patterns. This proactive and automated approach enhances the web application's defense against CSRF and XSS attacks, reducing the reliance on manual detection methods. By leveraging machine learning, organizations can fortify their security posture and respond promptly to evolving threats, contributing to a robust and resilient defense mechanism against these common web application vulnerabilities.

By leveraging machine learning, organizations can transition from traditional, rule-based security measures to a more adaptive and intelligent system. This approach reduces reliance on manual detection methods, enabling a quicker response to evolving threats. Additionally, machine learning provides the capacity to analyze vast amounts of information continuously, taking into consideration subtle and complex attack patterns that may go unnoticed by traditional security mechanisms. Moreover, the transition to machine learning-based detection reduces the reliance on manual rule creation and adjustment, streamlining the security process. The automated nature of ML-driven detection enhances the speed of response to emerging threats, contributing to a more robust and resilient security posture for web applications. To sum up, the integration of machine learning into CSRF and XSS attack detection represents a forward-thinking and dynamic strategy for enhancing web application security. This approach leverages the power of computational models to proactively identify and mitigate threats, adapting to the evolving nature of cyber risks in today's digital landscape. As organizations strive to stay ahead of malicious actors, the intelligent capabilities of machine learning provide a key advantage in fortifying the defenses of web applications against CSRF and XSS attacks.

## VIII. CONCLUSION AND FUTURE SCOPE

From the above presented algoXSSF framework and model, we have a security framework (of six steps - Recognize, Examine, Evaluate, Protect, Monitor, and Recoup) with machine learning algorithms embedded to combat malicious attacks, especially for XSRF and XSS attacks. The combination of emerging technology is urgently needed to combat the variants of cyber-attacks. The identification of cybercrime trends and attack patterns with step-wise improvement is possible within a blend of artificial intelligence, machine learning, and security framework, algoXSSF. The in-depth study apart from detection and analysis of two major cyberattacks (although there are many other variants of malware & cybercrimes ongoing in the online world) forms the solid foundational need of security model. Scripting attacks are possible from various ways including document object model, non-persistent assault, malicious payload, victim's browser plugins and others.

The essence and need of a security model apart from cyber security training, cyber awareness for every cyber-attack will be needed. In the future, we can even define flowchart including data flow diagrams for the proposed framework. Cyber-attacks on search engines and websites will definitely grow at a higher pace which needs more cybersecurity frameworks and cybersecurity models to counter them. As we delve into the "cookie" and "protocol" world, and online data apart from online users are growing every day, we will have to face new and more cyber-attacks in the future. Several vulnerable websites to inject scripts by attackers will be designed to trick and trap online users like it happened in the last year.

## REFERENCES

[1] Allioui, H., & Mourdi, Y. (2023). Exploring the Full Potentials of IoT for Better Financial Growth and Stability: A Comprehensive Survey. Sensors, 23(19), 8015.

[2] Khakimov, A. (2023). Impact of Cybercrime on individuals, businesses, and society. Best Journal of Innov in Science, Research & Dev, 213-216.

[3] Elluri, L., Mandalapu, V., Vyas, P., & Roy, N. (2023). Recent Advancements in Machine Learning for Cybercrime Prediction. Journal of Computer Information Systems, 1-15.

[4] Djenna, A., Barka, E., Benchikh, A., & Khadir, K. (2023). Unmasking Cybercrime with AI-Driven Cybersecurity Analytics. Sensors, 23(14), 6302.

[5] Ferreira, P., Caldeira, F., Martins, P., & Abbasi, M. (2023, February). Log4j Vulnerability. In International Conference on Information Technology & Systems (pp. 375-385). Cham: Springer International Publishing.

[6] Lei, Y., Lanson, J. P., Shue, C. A., & Wood, T. W. (2023, June). Attackers as Instructors: Using Container Isolation to Reduce Risk and Understand Vulnerabilities. In Int Conf on Detection of Intrusions & Malware, & Vul Asses (pp. 177-197). Cham: Spri Nature Switzerland.

[7] Madan, S., Savani, K., & Katsikeas, C. S. (2023). Privacy please: Power distance and people's responses to data breaches across countries. Journal of International Business Studies, 54(4), 731-754.

[8] Zagaris, B., & Mostaghimi, A. (2023). Cybercrime and Transnational Organized Crime. IELR, 39, 90.

[9] Amarendranath, M. K., Heena, S., Lalitha, B. N., & Umesh, D. S. (2023). Machine Learning for Web Vulnerability Detection: The Case of Cross-Site Request Forgery. Machine Learning, 52(4).

[10] Kshetri, N., Rahman, M.M., Sayeed, S.A., & Sultana, I. (2023). cryptoRAN: A review on cryptojacking and ransomware attacks w.r.t. banking industry - threats, challenges, & problems. *ArXiv, abs/2311.14783*. https://doi.org/10.48550/arXiv.2311.14783

[11] Kumari, S., Kumar Solanki, V., & Arokia Jesu Prabhu, L. (2023). Web Defenselessness Recognition Against Case of Cross Site Demand Fake. In Recent Dev in Electronics and Comm Systems (pp. 13-19). IOS Press.

[12] Yang, A., Lu, C., Li, J., Huang, X., Ji, T., Li, X., & Sheng, Y. (2023). Application of meta-learning in cyberspace security: A survey. Digital Communications and Networks, 9(1), 67-78.

[13] Noman, H. A., & Abu-Sharkh, O. M. (2023). Code Injection Attacks in Wireless-Based Internet of Things (IoT): A Comprehensive Review and Practical Implementations. Sensors, 23(13), 6067.

[14] Karimy, A. U., & Reddy, P. C. (2023). Securing the IoTs: A Study on ML-Based Solutions for IoT Security and Privacy Challenges

[15] Thajeel, I. K. T., Samsudin, K., Hashim, S. J., & Hashim, F. (2023). Machine and Deep Learning-based XSS Detection Approaches: A Systematic Literature Review. Journal of King Saud University-Computer and Information Sciences, 101628.

[16] Rodríguez, G. E., Torres, J. G., Flores, P., & Benavides, D. E. (2020). Cross-site scripting (XSS) attacks and mitigation: A survey. Computer Networks, 166, 106960.

[17] Malviya, V. K., Saurav, S., & Gupta, A. (2013, Dec). On security issues in web applications through cross site scripting (XSS). 20th asia-pacific software engineering conference (APSEC) (Vol. 1, pp. 583-588). IEEE.

[18] Krishnaraj, N., Madaan, C., Awasthi, S., Subramani, R., Avinash, H., & Mukim, S. (2023). Common vuln in real world web applications.

[19] Gedam, M. N., & Meshram, B. B. (2023). Proposed Secure Hypertext Model in Web Engineering. Journal of Web Engineering, 22(4), 575-596.

[20] Mireku Kwakye, M. (2022). Light-weight Privacy Infrastructure-A Blockchain-based Privacy-Preservation Platform for Data Storage and Query Processing.

[21] Mimura, M., & Yamasaki, T. (2022). Toward Automated Audit of Client-Side Vulnerability Against Cross-Site Scripting. In Advances on Broad-Band Wireless Computing, Communication and Applications: Proceedings of the 16th International Conference on Broad-Band Wireless Computing, Communication and Applications (BWCCA-2021) (pp. 148-157). Springer International Publishing.

[22] Liu, Z., Fang, Y., Huang, C., & Xu, Y. (2022). GAXSS: effective payload generation method to detect XSS vulnerabilities based on genetic algorithm. Security and Communication Networks, 2022, 1-15.

[23] Sadqi, Y., & Maleh, Y. (2022). A systematic review and taxonomy of web applications threats. Infn Sec Jour: A Global Persp, 31(1), 1-27.

[24] Aborujilah, A., Adamu, J., Shariff, S. M., & Long, Z. A. (2022, January). Descriptive Analysis of Built-in Security Features in Web Development Frameworks. In 2022 16th International Conf on Ubiquitous Information Management and Communication (IMCOM) (pp. 1-8). IEEE.

[25] Han, Y., Ji, X., Wang, Z., & Zhang, J. (2023, November). Systematic Analysis of Security and Vulnerabilities in Miniapps. In Proc of the 2023 ACM Workshop on Secure and Trustworthy Superapps (pp. 1-9).

[26] Kaur, J., Garg, U., & Bathla, G. (2023). Detection of XSS attacks using machine learning techniques: A review. AI Review, 1-45.

[27] Krishnan, M., Lim, Y., Perumal, S., & Palanisamy, G. (2022). Detection and defending the XSS attack using novel hybrid stacking ensemble learning-based DNN approach. Digital Communications and Networks.

[28] S. Calzavara, M. Conti, R. Focardi, A. Rabitti and G. Tolomei, "Mitch: A Machine Learning Approach to the Black-Box Detection of CSRF Vulnerabilities," 2019 IEEE EuroS&P, Stockholm, Sweden, 2019, pp. 528-543, doi: 10.1109/EuroSP.2019.00045.

[29] Krishnamoorthy, S. (2015, July 22). Cross Site Request Forgery Preventive Measures. Kongu. https://www.academia.edu/14316689/Cross_Site_Request_Forgery_Preventive_Measures

[30] What is a WAF? | web app firewall | Cloudflare. Available at: https://www.cloudflare.com/learning/ddos/glossary/web-application-firewall-waf/

[31] Jayawardana, H. D., Uyanahewa, M. I., Hapugala, V., & Thilakarathne, T. (2023, June 20). An Analysis of XSS Vulnerabilities and Prevention of XSS Attacks in Web Applications.

[32] What is cross-site scripting (XSS) and how to prevent it?: Web security academy (2023) What is cross-site scripting (XSS) and how to prevent it? https://portswigger.net/web-security/cross-site-scripting

[33] Singh, M., Singh, P. and Kumar, P. (2020) 'An analytical study on cross-site scripting', 2020 International Conference on Computer Science, Engineering and Applications (ICCSEA) [Preprint

[34] Bohara, R., Arjun, V. V., Jaiswal, J., Nikhil, M. R. S., Geetha, G., Pandey, B., & Raghav, U. R. (2023). A Survey On CROSS-SITE SCRIPTING.

[35] B. Gogoi, T. Ahmed, and H. K. Saikia, "Detection of XSS Attacks in Web Applications: A Machine Learning Approach," Int. J. Innov. Res. Comput. Sci. Technol., vol. 9(1),1–10, 2021, 10.21276/ijircst.2021.9.1.1.

[36] Zafar, M.F., Rawat, N., Mishra, R., Shekhar Pandey, P., Kshetri, N. (2023). Uncovering Deception: A Study on Machine Learning Techniques for Fake News Detection. ICICCT 2023. Lect Notes in Nw and Sys, vol 757. Sprin, https://doi.org/10.1007/978-981-99-5166-6_56

[37] Wang, Q., Yang, H., Wu, G., Choo, K. K. R., Zhang, Z., Miao, G., & Ren, Y. (2022). Black-box adversarial attacks on XSS attack detection model. Computers & Security, 113, 102554.

[38] Kshetri, N., Hutson, J., & Revathy, G. (2023). healthAIChain: Improving security and safety using Blockchain Technology applications in AI-based healthcare systems. *ArXiv, abs/2311.00842*. https://doi.org/10.48550/arXiv.2311.00842

[39] Habibie, M. I., & Nurda, N. (2022, August). Performance Analysis and Classification using Naive bayes and Logistic Regression on Big Data. In 2022 1st International Conference on Smart Technology, Applied Informatics, and Engineering (APICS) (pp. 48-52). IEEE.

[40] Kascheev, S., & Olenchikova, T. (2020). The detecting cross-site scripting using machine learning methods. In GloSIC, pp. 265-270, IEEE

[41] Banerjee, P., Chattopadhyay, T., & Chattopadhyay, A. K. (2023). Comparison among different Clustering and Classification Techniques: Astronomical data-dependent study. New Astronomy, 100, 101973.

[42] Roy, P., Kumar, R., Rani, P., & Joy, T. S. (2022, December). XSS: Cross-site Scripting Attack Detection by ML Classifiers. In 11$^{th}$ Int Conf on Sys Modeling & Adv in Research Trends (SMART), 1535-1539, IEEE.

[43] Hiremath, S., Shetty, E., Prakash, A. J., Sahoo, S. P., Patro, K. K., Rajesh, K. N., & Pławiak, P. (2023). A New Approach to Data Analysis Using ML for Cybersecurity. Big Data & Cognitive Comp, 7(4), 176.

[44] Kshetri, N., Vasudha, & Hoxha, D. (2023). knowCC: Knowledge, awareness of computer & cyber ethics between CS/non-CS university students. *abs/2310.12684*. https://doi.org/10.48550/arXiv.2310.12684

[45] Shahid, M. (2023). Machine Learning for Detection and Mitigation of Web Vulnerabilities and Web Attacks. arXiv preprint arXiv:2304.14451.

[46] Lei, L., Chen, M., He, C., & Li, D. (2020, October). XSS detection technology based on LSTM-attention. In 2020 5th International Conf on Control, Robotics and Cybernetics (CRC) (pp. 175-180). IEEE.

[47] Lente, C., Hirata Jr, R., & Batista, D. M. (2021, October). An Improved Tool for Detection of XSS Attacks by Combining CNN with LSTM. In Anais Estendidos do XXI Simpósio Brasileiro em Segurança da Informação e de Sistemas Computacionais (pp. 1-8). SBC.

[48] G. Bano and B. Mastoi, "Framework for Monitoring and Detection of DDOS Attacks using ML Algorithms," no. December 2022

[49] S. Kumara, V. Kumar, and S. Phd, "Web Vulnerability Detection: the Case of Cross-Site Request Forgery," Ijarst.in, vol. 12, no. 01, p. 1, 2022

[50] Naresh Kshetri, Chandra Sekhar Bhushal, Purnendu Shekhar Pandey and Vasudha, "BCT-CS: Blockchain Technology Applications for Cyber Defense and Cybersecurity: A Survey and Solutions" Int. Journal of Advanced Computer Science and Applications (IJACSA), 13(11), 2022. https://dx.doi.org/10.14569/IJACSA.2022.0131140